# Synthetic social data: trials and tribulations

Guido Ivetta[1]†, Laura Moradbakhti[2]
and Rafael A. Calvo[2]†

guidoivetta@unc.edu.ar
†These authors contributed equally to this work.

[1] FAMAF, Universidad de Cordoba, Argentina.
[2] Dyson School of Design Engineering, Imperial College London.

**Abstract**
Large Language Models are being used in conversational agents that simulate human conversations and generate social studies data. While concerns about the models' biases have been raised and discussed in the literature, much about the data generated is still unknown.
In this study we explore the statistical representation of social values across four countries (UK, Argentina, USA and China) for six LLMs, with equal representation for open and closed weights. By comparing machine-generated outputs with actual human survey data, we assess whether algorithmic biases in LLMs outweigh the biases inherent in real- world sampling, including demographic and response biases. Our findings suggest that, despite the logistical and financial constraints of human surveys, even a small, skewed sample of real respondents may provide more reliable insights than synthetic data produced by LLMs. These results highlight the limitations of using AI-generated text for social research and emphasize the continued importance of empirical human data collection.

## 1. Introduction

Understanding public opinion is essential for social scientists seeking to analyse societal trends, cultural shifts, and policy implications. Traditionally, large-scale surveys, such as the World Values Survey (WVS), have provided valuable insights into public attitudes on a range of issues. However, these surveys require extensive planning, substantial funding, and significant time investments to ensure representative and reliable results.

With the rapid advancement of artificial intelligence, particularly Large Language Models (LLMs), researchers have begun exploring the potential of these models as a cost-effective alternative to traditional surveys. LLMs are trained on vast amounts of text data, capturing patterns in human language and potentially reflecting social attitudes. This raises an intriguing question: Can LLMs generate synthetic data that approximates real human responses? If so, could they be used as a preliminary or even a supplementary tool for social research?

Despite growing interest in this approach, the accuracy and validity of LLM-generated data remain uncertain. While some argue that LLMs can simulate public opinion by mirroring linguistic trends, concerns persist about machine-generated biases that could distort results. Specifically, LLMs inherit biases from their training data and algorithmic design, which may introduce systematic errors not present in traditional sampling methods.

## 2. Related Work

The use of Large Language Models (LLMs) as potential tools for social science research has gained significant attention in recent years. Researchers have explored whether these models can simulate public opinion and serve as substitutes for human survey respondents. However, concerns regarding bias, accuracy, and representativeness remain central to this discussion.

Several studies have investigated the feasibility of using LLMs to simulate human attitudes and opinions. (Qu & Wang, 2024) examined ChatGPT's ability to reflect public opinion based on socio-demographic data from the World Values Survey (WVS). Their findings indicate that while LLMs demonstrate some ability to approximate human responses, they exhibit significant biases across demographic and thematic dimensions. Specifically, models performed better in Western, English-speaking, and developed nations while struggling with responses from non-Western and less-developed countries. Furthermore, demographic biases emerged in relation to gender, ethnicity, age, education, and social class, raising concerns about the equitable application of LLMs in public opinion research.

(Boelaert et al., 2024) similarly examined the viability of LLMs in survey research and introduced the concept of "machine bias." Their study demonstrated that LLMs do not systematically favor demographic groups in a predictable manner. Instead, the biases embedded in LLM-generated responses vary randomly across different topics. They argue that LLMs display a strong bias with low variance, meaning that their answers tend to cluster around certain patterns rather than mimicking the natural variability found in human survey responses. As a result, they conclude that LLMs cannot replace human subjects for opinion or attitudinal research.

Other scholars have also weighed in on this debate. (Argyle et al., 2023) found that ChatGPT could approximate political attitudes by generating responses similar to those of different socio-demographic subgroups. They introduced the concept of "silicon sampling," wherein LLMs generate synthetic survey responses based on human-like demographic conditioning. While their study suggested a promising level of alignment between LLM-generated and human responses, critics have pointed out that these models still struggle with more complex, nuanced questions, particularly those requiring deep cultural or historical understanding (Bisbee et al., 2024).

## 3. Ethics and biases

The issue of bias in LLMs is well documented. Studies have shown that LLMs often reflect the cultural and ideological biases present in their training data (Bender et al., 2021); (Santurkar et al., n.d.); (Fort et al., 2024). (Boelaert et al., 2024) emphasizes that, rather than consistently favouring one particular group, LLMs generate unpredictable biases depending on the topic at hand. This contradicts the "social bias hypothesis," which assumes that LLMs primarily reflect the perspectives of dominant social groups. Instead, the findings suggest that LLMs' outputs fluctuate in ways that make them unreliable for systematic social science research.

Additionally, research highlights the limitations of LLMs in addressing complex topics that require deep contextual knowledge. For example, (Motoki et al., 2024) found that ChatGPT exhibited political biases that leaned toward liberal perspectives, while (von der Heyde et al., 2025) noted inconsistencies in responses to political questions in Germany. These findings reinforce concerns that LLMs may not be suitable for unbiased public opinion research.

Despite the promise of LLMs as cost-effective alternatives to traditional surveys, studies continue to highlight the strengths of human-based survey methods. While declining response rates and high costs pose challenges (Dutwin & Buskirk, 2021) traditional surveys still provide a level of depth, nuance, and representativeness that LLMs cannot currently match. In this paper, we argue that even small, non-representative human samples often yielded more accurate insights than LLM-generated data, further questioning the reliability of synthetic responses. Unlike human survey respondents, LLMs generate responses based on patterns learned from vast amounts of text data. This introduces several layers of potential bias:

1. Training Data Bias: LLMs are trained on internet-based corpora that may overrepresent certain demographics (e.g., English-speaking, Western, and urban populations) while underrepresenting others (e.g., rural, low-income, or non-digital- native groups). Studies have shown that models like GPT-4 perform better on Western-centric datasets than on those from non-Western contexts (Qu & Wang, 2024).
2. Algorithmic Bias: The way LLMs process and generate responses can introduce systematic distortions. Prior research (Boelaert et al., 2024) has demonstrated that LLMs tend to produce answers with low variance, meaning they generate overly homogeneous responses that do not fully capture the diversity of human opinion.
3. Synthetic Representation Bias: Although efforts have been made to engineer "synthetic respondents" that reflect diverse perspectives, these representations may not be truly independent but rather extrapolations based on the biases of the training data. For instance, while an LLM can be prompted to respond as a "working-class woman in China," its response is still influenced by the overall distribution of texts it has been trained on, rather than actual survey data from that demographic group.

## 4. Dataset: World Values Survey (WVS) - Wave 7

The World Values Survey (WVS) is a comprehensive global research project that explores people's values and beliefs, how they change over time, and their social and political impact. Since its inception in 1981, the WVS has conducted seven waves of surveys, with the seventh wave (2017-2020) encompassing data from 77 countries, including Argentina, the United States, the United Kingdom, and China. This data is publicly available and can be accessed through the official WVS website

### 4.1 Sample Sizes and Methodology

In Wave 7, the WVS employed rigorous sampling techniques to ensure nationally representative data. Each participating country aimed for a minimum sample size of 1,000 respondents aged 18 and above. The survey utilized stratified random sampling methods, considering factors such as region, urbanization level, and demographic characteristics to accurately reflect each nation's population.

### 4.2 Questionnaire Structure

The WVS Wave 7 questionnaire is organized into 14 thematic sections, covering a wide array of topics, we selected questions related to economic views and wellbeing. For the purposes of this study, questions specifically related to economic views and wellbeing were selected from the questionnaire. These questions included:
1. Household Economic Hardship: This included inquiries about experiences such as lacking sufficient food, feeling unsafe from crime at home, being without needed medicine or medical treatment, experiencing a lack of cash income, and not having safe shelter.
2. Household & Living Standards: Questions in this area assessed satisfaction with one's current financial situation and a rating of current living standards compared to parents' at the same age.
3. Views on Society & Economy: This section explored perspectives on the balance between income equality and incentives for effort, private versus government ownership, government provision versus individual responsibility, the impact of competition, and whether success is attributed more to hard work or luck/connections.
4. Democratic Values: This category included questions on the essentiality of government policies like taxing the rich to subsidize the poor, state aid for unemployment, civil rights protecting from state oppression, and the state making people's incomes equal as components of democracy.

## 5. Methods

In this section, we describe the methodology for our experiments. We first examine key demographic variables commonly employed in social science, providing a foundation for systematic comparison. We then detail the experimental design, including procedures for both data generation and analysis.

### 5.1 Synthetic representation

To assess bias in Large Language Models (LLMs) when simulating public opinion, we examine five demographic variables—Age, Social Class, Education Level, Sex, and Country of Origin—selected for their established relevance in structuring attitudes and behaviours and for their comparability across surveys. We pair these with four countries—Argentina, the United States, the United Kingdom, and China—chosen to span major world regions and distinct cultural–linguistic and socio-economic contexts (Latin America, North America, Europe, and East Asia), enabling tests across both Western and non-Western populations. The design is constrained by computation: exhaustively covering a larger set of variables and countries would create a combinatorial space that exceeds available resources. Within these limits, our selection prioritizes coverage of diverse cultures and populations while preserving tractable, replicable analyses.

Our methodology for generating synthetic responses was adapted from the interview-style simulation approach proposed by (Qu & Wang, 2024). For each unique combination of demographic variables, we constructed a detailed persona prompt by translating the categorical data for age, social class, education, sex, and country into descriptive, natural-language sentences. These sentences were then integrated into a single prompt that instructed the LLM to adopt the specified identity before answering a survey question. A typical prompt was structured as follows:

> *"Assuming you are a [Age]-year-old [Sex] from [Country] with a [Education Level] education who identifies as [Social Class], how would you answer the following question: [WVS Question Text and Response Options]?"*

Following the practices outlined by (Qu & Wang, 2024) to preserve cultural and linguistic context, we utilized a multilingual approach. Prompts for Argentina and China were provided in Spanish and Mandarin, respectively, using the official translations from the WVS questionnaires. For the United States and the United Kingdom, prompts were administered in English.

### 5.2 Hypothesis and Research Approach

The use of Large Language Models (LLMs) to simulate survey respondents presents a methodological trade-off. While this approach eliminates challenges of sample size and recruitment, it introduces a systematic machine bias rooted in the model's training data and architecture. Conversely, traditional surveys are constrained by practical limitations, making them susceptible to sampling errors, particularly when samples are small or fail to represent the target population. This study assesses the relative magnitude of these two error types to determine which approach yields a more faithful estimate of population characteristics.

Our central hypothesis is that the bias from LLM-generated responses is more significant than the selection bias found in imperfect, yet conventional, human surveys. We posit that across a range of demographic variables and their intersections, even a small or unrepresentative human sample will align more closely with population benchmarks than synthetic data will. Our research approach is designed to test this hypothesis by directly comparing the outputs of both methods against established ground-truth data.

Our methodology unfolds in three stages. First, we establish a ground-truth benchmark by analyzing large-scale human survey data, in particular the World Values Survey (WVS), to determine the expected response distributions for our selected demographic variables. Second, we generate a corresponding set of synthetic responses using an LLM, conditioning the model to simulate the same demographic profiles. The core of our analysis is the final stage, where we quantify the discrepancies between the human-generated and machine-generated data. By employing statistical measures such as ANOVA and Tukey's HSD, we directly compare the variance and distribution of responses. This approach allows us to rigorously test whether LLMs introduce systematic distortions that exceed the known errors of traditional survey sampling, thereby addressing our central hypothesis about the comparative magnitude of machine bias versus sampling bias.

## 6. Results

Our analysis quantitatively compared survey responses from six Large Language Models (LLMs) against human data from the World Values Survey (WVS). The WVS questions featured response scales of 1-to-4 or 1-to-10 points. The initial analysis compared the aggregated average response of each model against the aggregated average human response across four countries (United Kingdom, Argentina, the United States, and China) for 15 selected questions. Figure 1 presents the outcomes of this comparison. The average human response for each question is shown as a black dashed line, serving as the benchmark. The average responses for the six LLMs are plotted as coloured markers. Fifteen one-way ANOVAs was conducted: one for each question. This was followed by Tukey's Honestly Significant Difference (HSD) post-hoc test to compare each model's mean response to the human mean. A marker displayed as an 'x' denotes a statistically significant difference ($p \leq 0.05$), while a '✓' indicates the difference was not statistically significant ($p > 0.05$). Of the 90 model-question comparisons, 94.4% of the LLM-generated responses were statistically different ($p \leq 0.05$) from the human benchmark. This foundational finding establishes a general misalignment between the outputs of the LLMs and the human survey data.

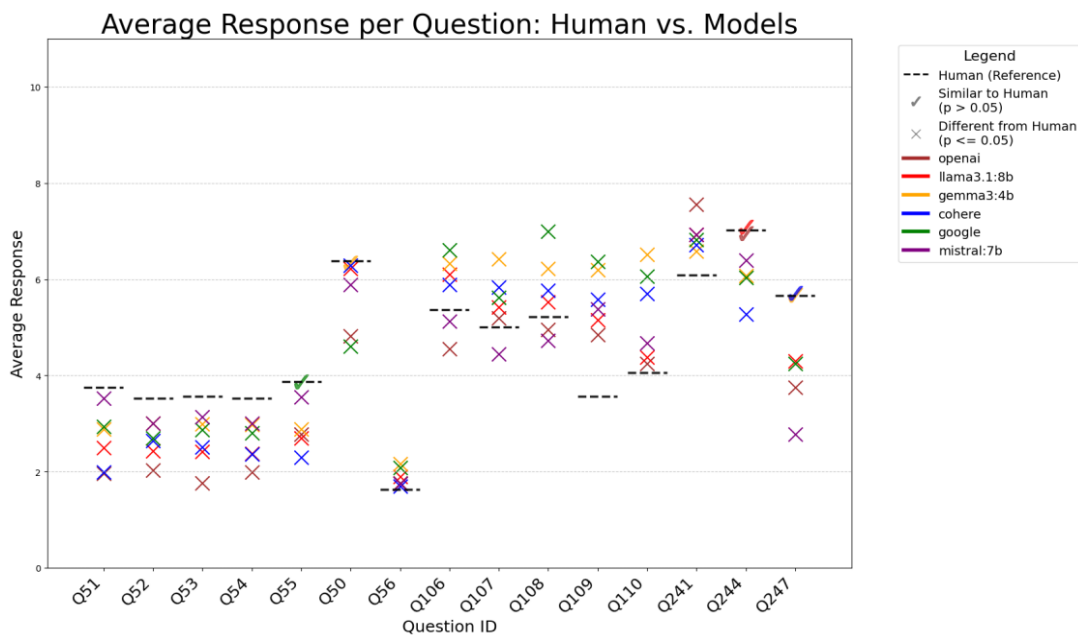

*Figure 1.*

*This plot compares the average responses of six AI models against the average human response (black dashed line) for several questions. For each question, a one-way ANOVA confirmed a statistically significant difference among the full set of responses (i.e., human and all models). The markers shown here visualize the subsequent post-hoc tests comparing each model individually to the human reference: an 'x' denotes a statistically significant difference ($p \leq 0.05$), while a '✓'*

*denotes no significant difference (p>0.05). The overwhelming prevalence of 'x' markers illustrates a significant divergence, with 95% of the AI models' responses being statistically different from the human benchmark.*

Given this divergence, we next investigated the central hypothesis: whether the sampling bias in small human samples is less significant than the machine bias inherent in LLM-generated data. We assessed this by analysing the performance of human samples of varying sizes against the synthetic data. Figure 2 illustrates the percentage of cases where the average of a randomly drawn human sample was closer to the true population mean than the average of the synthetic data was. The results show a clear relationship between sample size and reliability. A single human respondent (N=1) was more accurate than the synthetic data in 47.5% of cases, but this figure rises to 60.8% with just two respondents (N=2). The performance improves rapidly as sample size increases, surpassing 86% at a sample size of 16 and reaching 97.6% at 128 respondents.

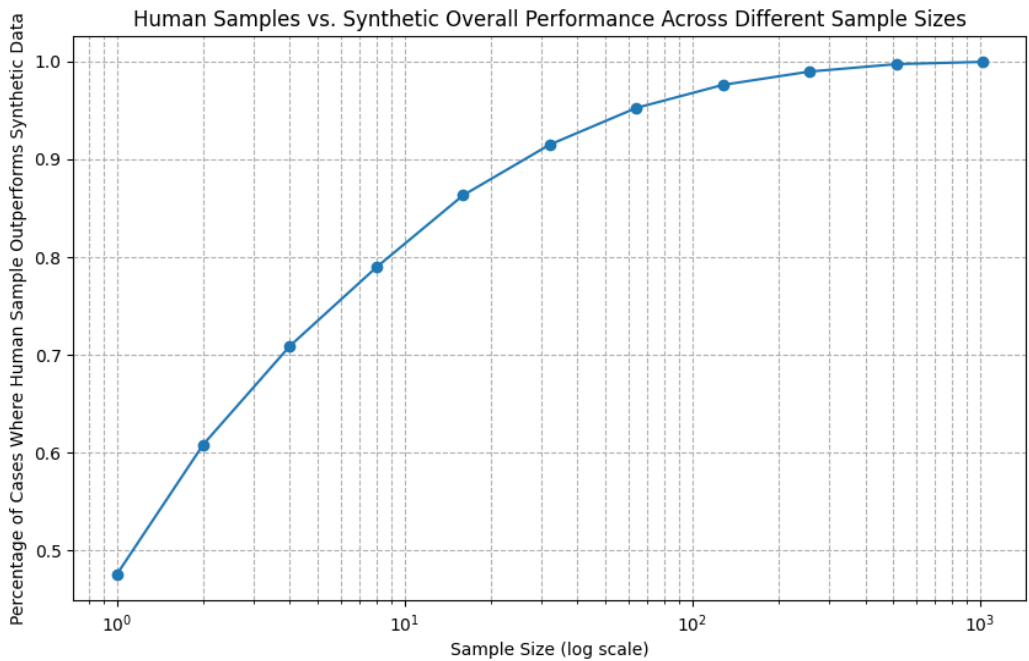

*Figure 2.*

*Even very small and potentially biased human samples outperform synthetic data, and as sample sizes increase, performance improves rapidly. By the time just 16 people are sampled, human data is already significantly more reliable than synthetic alternatives. This reinforces the idea that even a small, skewed human dataset is preferable to relying entirely on machine-generated responses.*

To further examine the impact of sampling bias, we analyzed the performance of specific, demographically defined subgroups within the human data. Figure 3 plots the "Human Closer Ratio" for every demographic combination available in the dataset (e.g., by age, social class, education, sex, and country). Each point represents a subgroup, and its position on the y-axis indicates how often its members' average response was more accurate than the synthetic data. The red dashed line at 50% marks the threshold above which a human subgroup outperforms the machine. The vast majority of subgroups, including those with very small and specific populations such as "Upper Class-Primary education" (n=3) and "Post secondary non-tertiary education-Male" (n=3), fall above this 50% line. This demonstrates that even small or highly skewed human samples are, on average, more representative of the population's views than the data generated by the LLMs.

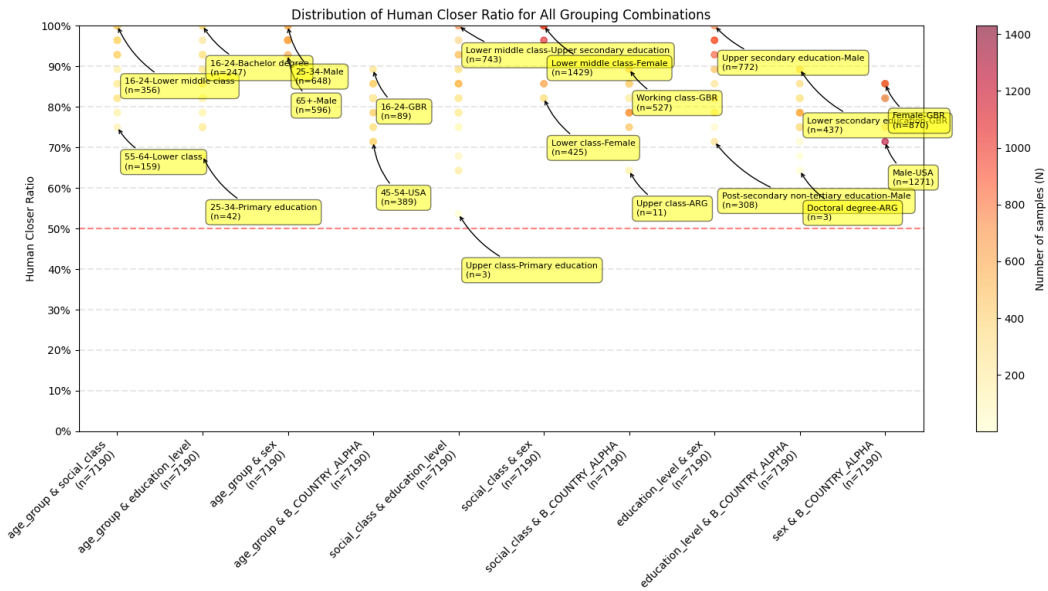

*Figure 3.*

*One of the key takeaways from this graph is that even small, biased human samples outperform the machine in most cases. For example, some subgroups with as few as 3 to 42 individuals still achieve significantly better accuracy than machine-generated responses. The red dashed line at 50% highlights that all subgroups above this threshold indicate human superiority. While larger groups generally show strong performance, smaller groups are still highly effective, demonstrating that even a highly skewed or limited human sample can yield better results than synthetic data. This reinforces the conclusion that sampling bias is less damaging than machine bias when trying to approximate real-world distributions.*

Collectively, these results suggest that while LLM-generated data consistently differs from human averages, even small or demographically skewed human samples provide a more reliable estimate of population responses than the synthetic data generated by the tested models.

## 7. Discussion

This study compared Large Language Model (LLM) outputs with human survey data to determine which is a more reliable source for social science research. The results showed that LLM responses were consistently different from human responses and that even small samples of human data were more accurate than the data generated by the six models tested. This section discusses the meaning of these findings, their implications for research, the challenges of AI in data collection, and areas for future work.

### 7.1. Interpretation of Findings

Our study was designed to compare two potential sources of error: the inherent bias in LLMs ("machine bias") and the bias that comes from using small or non-representative samples of people ("sampling bias"). The results indicate that machine bias, as it exists in the current models, is a more significant problem than sampling bias. The fact that all six LLMs produced outputs that were statistically different from the human averages on most questions (Figure 1) shows that these models do not accurately reflect human values.

Furthermore, the analyses in Figures 2 and 3 showed that human data becomes more reliable than synthetic data at very small sample sizes. A sample of just two people was more accurate than the models over 60% of the time. This suggests that even with the limitations of survey sampling, collecting data from a small number of actual people provides a better estimate of a population's views than relying on current LLMs. This is likely because LLMs, despite being trained on vast amounts of human text, do not capture the underlying reasoning or lived experience that informs human values.

## 7.2. Implications for Social Science Research

The findings have two main implications for researchers. First, using LLMs as direct substitutes for human survey participants is not advisable currently. Doing so would likely produce poor-quality data that leads to incorrect conclusions about society. The biases in LLMs are not random but systematic, which could create a false sense of confidence in misleading results.

Second, the clear distinction between "human data" and "synthetic data" is becoming less certain. Research from Zhang (2025) indicates that a substantial number of people (around 34%) use AI tools to help answer survey questions. This creates a new methodological challenge: researchers may be unknowingly collecting AI-influenced data from participants. This can corrupt datasets by introducing the very machine biases our study identified, making it harder to understand true human opinions.

## 7.3. Human-AI Interaction and Value Alignment

As people interact more with AI agents, they may form assumptions about the AI's values. For instance, a user might believe an AI is objective, has no values, or shares the user's own values. Our results (Figure 1) show these assumptions are incorrect; the tested AIs have distinct and unpredictable biases.

This mismatch between user perception and the AI's actual behavior is a concern. If a person believes an AI is a neutral source of information, they may be influenced by its biased responses. This could subtly shift the person's own stated opinions, especially on complex topics. This highlights the need for transparency about the limitations and inherent biases of AI systems that people interact with.

## 8. Conclusions

This study was motivated by a central question in contemporary social science: Can Large Language Models (LLMs) serve as reliable substitutes for human survey respondents? Proponents of this approach highlight its potential to overcome the logistical and financial barriers of traditional surveys, offering a seemingly endless supply of synthetic data. However, this raises a critical trade-off between the known challenges of human sampling bias and the unknown dimensions of machine bias. Our research sought to empirically test this trade-off by comparing responses from six prominent LLMs against authentic human data from the World Values Survey across four diverse countries.

Our findings deliver a clear verdict. First, we identified a fundamental and widespread divergence between machine and human responses, with over 94% of LLM-generated answers being statistically different from the human benchmark. This indicates that current models do not accurately replicate aggregate human social values. Second, and more significantly, our analysis demonstrated the consistent superiority of authentic human data, even when samples were small and potentially unrepresentative. A random sample of just two human participants was more reliable than the synthetic data in most cases, an advantage that held even for niche demographic subgroups. This provides strong evidence that the systematic and opaque nature of machine bias is a more significant threat to validity than the sampling bias inherent in small-scale human surveys.

The implications of this research are twofold. For social scientists, it serves as a strong caution against the premature adoption of LLMs as proxies for human subjects, as doing so risks building theories on flawed and unrepresentative data. For the broader fields of AI ethics and development, our work highlights critical challenges, including the methodological problem of AI-generated text contaminating human datasets and the societal risk of deploying biased agents that can subtly influence users. While the promise of AI in research is vast, this study reaffirms that for the purpose of understanding human values, beliefs, and attitudes, there is currently no substitute for data gathered from humans themselves. The continued, careful collection of authentic human data, even with its imperfections, remains essential for sound social science.

## Limitations

Our analysis, based on six models, 15 questions, and four countries, should be understood as an illustrative snapshot rather than an exhaustive census of the rapidly evolving AI ecosystem. Methodologically, our prompting strategy represents one of many possible approaches that can influence model outputs, and our aggregation of country-level data demonstrates a general trend at the cost of masking important local variations. Future work should therefore expand this comparative framework to more models and cultural contexts, conduct disaggregated analyses, and systematically investigate the impact of prompt engineering on model representativeness.

Beyond the scope of this study, we caution against the simple assumption that future models will inherently become better proxies for human opinion. The process of AI "alignment" may sanitize responses according to a narrow, often Western-centric, set of values, paradoxically making models less representative of diverse global viewpoints. This suggests machine bias is a persistent challenge of value encoding, not a temporary technical flaw that will be easily fixed. Consequently, two critical research priorities emerge for the field: first, developing robust methods to detect synthetic text in survey data to ensure research integrity, and second, conducting empirical studies to measure how interaction with biased AI agents may influence human attitudes and beliefs.

## Ethical Considerations

The use of Large Language Models in social science introduces complex ethical challenges that warrant careful consideration. The research presented in this paper was conducted following established ethical guidelines, utilizing the publicly available and anonymized World Values Survey (WVS) dataset, ensuring no harm to the original human participants. Our methods involving the AI models were transparently reported to allow for scrutiny. However, the findings themselves—that LLMs do not accurately represent human social values—raise significant ethical implications for the development and deployment of this technology.

If LLMs were to function as intended, providing a fast, cheap, and accurate simulation of human opinion, the primary beneficiaries would be researchers, non-profits, and public bodies with limited resources. This could democratize access to public opinion data, allowing more organizations to conduct research. However, our findings suggest the technology does not function as intended. The primary harm, therefore, stems from the technology's failure to perform as promised. When an LLM produces flawed data that is mistaken for genuine human opinion, the consequences can be severe. Policymakers, a key stakeholder group, might design ineffective social programs or legislation based on a distorted understanding of public needs. Social scientists could build incorrect theories about human behavior. In this failure mode, the group ultimately harmed is the general public, whose actual values and needs are misrepresented and ignored.

The mechanisms for this failure are the biases embedded within the models, which originate from their training data and algorithmic design. These datasets often overrepresent Western, English-speaking, and internet-connected populations, while underrepresenting other groups. This can lead to specific harms when the technology is misused. For example, a company might use an LLM to create a "synthetic focus group" to test a new health product. If the model's biases cause it to poorly represent the views of elderly or rural populations, the company might wrongly conclude the product is suitable for everyone. This could lead to a product that fails to meet the needs of those very groups, wasting resources and potentially causing harm if the product is inaccessible or ineffective for them.

These harms are not distributed equally across society; they fall disproportionately on populations that already experience marginalization. Vulnerable stakeholders—including ethnic and linguistic minorities, low-income communities, and people in the Global South—are often the least represented in training data. Consequently, their perspectives are the most likely to be distorted or silenced by LLM-generated data. This creates a dangerous feedback loop where a lack of representation in technology leads to policies and products that further entrench societal inequalities. The ethical imperative is to recognize that the stakeholders most at risk are those with the least power to influence the technology's development.

Finally, a broader ethical issue concerns the potential for these systems to covertly influence users. When people interact with an AI, they may assume it is a neutral source of information. As our research shows, these models are not neutral. This creates a risk that the AI's biased outputs could subtly shape a user's beliefs without their awareness, raising concerns about manipulation and personal autonomy. This underscores the need for greater transparency from developers about model limitations and a greater sense of responsibility from researchers and practitioners. The potential benefits of synthetic data must be carefully weighed against the clear and present ethical risks of misrepresentation, bias amplification, and the disproportionate harm to vulnerable communities.

## Acknowledgements

This work used computational resources from CCAD – Universidad Nacional de Córdoba (https://ccad.unc.edu.ar/), which are part of SNCAD – MinCyT, República Argentina. It was also supported by the computing power of Nodo de Cómputo IA, from Ministerio de Ciencia y Tecnología de la Provincia de Córdoba in San Francisco - Córdoba, Argentina.

# Appendix

| Group | Code | Question |
|---|---|---|
| **Household Economic Hardship** | Q51 | How often in the last 12 months have you or your family gone without enough food to eat? *(1='Often', 4='Never')* |
| | Q52 | How often in the last 12 months have you or your family felt unsafe from crime in your home? *(1='Often', 4='Never')* |
| | Q53 | How often in the last 12 months have you or your family gone without medicine or medical treatment that you needed? *(1='Often', 4='Never')* |
| | Q54 | How often in the last 12 months have you or your family gone without a cash income? *(1='Often', 4='Never')* |
| | Q55 | How often in the last 12 months have you or your family gone without a safe shelter over your head? *(1='Often', 4='Never')* |
| **Household & Living Standards** | Q50 | How would you rate your satisfaction regarding your current financial situation? *(1='Dissatisfied', 10='Satisfied')* |
| | Q56 | How would you rate your current standard of living compared to your parents' at your age? *(1='Better off', 2='Worse off', 3='About the same')* |
| **Views on Society & Economy** | Q106 | Where do you fall between: *Incomes should be more equal (1)* and *Greater incentives for effort (10)*? |
| | Q107 | Where do you fall between: *Private ownership should increase (1)* and *Government ownership should increase (10)*? |
| | Q108 | Where do you fall between: *Government should ensure provision (1)* and *People should provide for themselves (10)*? |
| | Q109 | Where do you fall between: *Competition is good (1)* and *Competition is harmful (10)*? |

| | Q110 | Where do you fall between: *Hard work brings success (1)* and *Success depends on luck/connections (10)*? |
|---|---|---|
| **Democratic Values** | Q241 | How essential is this to democracy: *Governments tax the rich and subsidize the poor*? *(1='Not essential', 10='Essential')* |
| | Q244 | How essential is this to democracy: *People receive state aid for unemployment*? *(1='Not essential', 10='Essential')* |
| | Q247 | How essential is this to democracy: *The state makes people's incomes equal*? *(1='Not essential', 10='Essential')* |